\documentclass[11pt,review,1p,number,sort&compress,a4paper]{elsarticle}

\usepackage{epsfig,color}
\usepackage{graphicx}
\usepackage{url}
\usepackage{soul}

\newcommand{\eqref}[1]{(\ref{#1})}

\newcommand{\3}{$_{3}$}

\newcommand{\cm}{cm$^{-1}$}
\newcommand{\p}{^\prime}
\newcommand{\pp}{^{\prime\prime}}

\newcommand{\nhhh}{$^{15}$NH$_3$}

\newcommand{\ai}{\textit{ab initio}}

\title{A theoretical room-temperature line list for $^{15}$NH$_3$ }

\author{Sergei N. Yurchenko\corref{cor1}}
\address{Department of Physics and Astronomy, University College London, London, WC1E 6BT, UK}

\cortext[cor1]{Corresponding author. Tel: +442076790172; Fax: +442076797145; E-mail: s.yurchenko@ucl.ac.uk}

\date{Accepted XXXX. Received XXXX; in original form XXXX}

\begin{document}

\begin{abstract}

A new room temperature line list for $^{15}$NH$_3$ is presented. This line
list comprised of transition frequencies and Einstein coefficients has been
generated using the `spectroscopic' potential energy surface NH3-Y2010 and an
\textit{ab initio} dipole moment surface. The $^{15}$NH$_3$ line list is
based on the  same computational procedure used for the line list for
$^{14}$NH$_3$  BYTe reported recently  and should be as  accurate.
Comparisons with experimental frequencies and intensities are presented.
The synthetic spectra show excellent agreement with experimental spectra.

\end{abstract}

\maketitle

\section{Introduction}


Although the abundance of $^{15}$N is 450 times lower than that of $^{14}$N
\cite{04AbLeOw.15NH3}, $^{15}$NH\3\ is an important astrophysical molecule. It is a maser
source detected in interstellar molecular clouds \citep{91ShWaMa.15NH3} and is also a
tracer of the N$^{15}$/N$^{14}$ isotopic ratio in interstellar medium
\citep{02ChRoxx.15NH3,10LiWoGe.15NH3,09GeMaBi.15NH3},  planetary
\citep{00FoLeBe.15NH3,04FoIrPa.15NH3} and Earth \citep{90HaShxx.15NH3} atmospheres,
meteorites \citep{12PiWixx.15NH3}, comets \citep{11MuChxx.15NH3}, important as a probe of
chemical processes in the astrophysical environment, of planetary atmospheric and
formation processes etc. Very recently \citet{14FlGrOr.15NH3} used the $^{14}$NH\3\ and
$^{15}$NH\3\ spectral features to study the N$^{15}$/N$^{14}$  ratio for Jupiter and
Saturn.










Experimentally the ro-vibrational spectra of $^{15}$NH\3\ have been studied in
a large number of works, including rotation-inversion spectrum
\citep{70ShShxx.15NH3,80CaTrVe.15NH3,80Saxxxx.15NH3,91FuCaDi.15NH3,94ScReFu.15NH3,94UrKlYa.15NH3},
fundamental bands
\citep{80Coxxxx.15NH3,83JoPaDc.15NH3,83UrPaBe.15NH3,84UrDcRa.15NH3,85UrDcRa.15NH3,85DcUrRa.15NH3,86IwUeNa.15NH3,86UrDcMa.15NH3,00AnJoLe.15NH3,11FuNiSp.15NH3},
overtone bands
\citep{82DiFuTr.15NH3,82SaHaAm.15NH3,90FuBaxx.15NH3,91MoNaSh.15NH3,06LeLiLi.15NH3,08LeLiXu.15NH3},
hot bands \citep{77KaKrPa.15NH3,85UrMiRa.15NH3,86SaScxx.15NH3,87Dcxxxx.15NH3},
and intensity measurements
\citep{81VaWyxx.15NH3,90DeRaPr.15NH3,95SiMaWh.15NH3,11LiPfEn.15NH3}. Some of
these data are now collected in the HITRAN database \cite{HITRAN2012}. The
electric dipole moment was experimentally studied by \citet{77OrOkxx.15NH3} and
\citet{81DiTrVe.15NH3} using the Stark spectroscopy.  The ground state energies
were reported by \citet{85UrDcRa.15NH3}. Very recently a VECSEL laser source
study of the 2.3~$\mu$m region of \nhhh\ was presented by
\citet{14CeHoVe.15NH3}  and a tentative assignment of new \nhhh\ lines in the
1.51~$\mu$m region was suggested by \citet{14FoGoHe.NH3}.

\citet{11HuScLe2.NH3} presented theoretical ro-vibrational energies of
$^{15}$NH\3\ computed variationally using an empirical PES HSL-2 for
$J=0\ldots\ 6$. These energies helped them to reassign and correct a number of
transitions in HITRAN. An extensive hot line list BYTe for $^{14}$NH\3\ was
recently generated \citep{11YuBaTe.NH3} using the TROVE
approach~\cite{07YuThJe.method}. Containing 1.1 billion transitions BYTe was
designed to be applicable for temperatures up to 1500~K. It has proven to be
useful for astrophysical  and spectroscopic applications (see, for example,
Refs. \cite{11BeTiKi.exo,11LuTiBu.dwarf,14CeHoVe.15NH3}). In this work we build
a room temperature line list for the N$^{15}$ isotopologue of ammonia using the
same computational approach based on the `spectroscopic' potential energy
surface (PES) NH3-Y2010 \cite{11YuBaTe.method} and the \ai\ dipole moment
surface (DMS) from Ref.~\cite{09YuBaYa.NH3}. The highest $J$ considered in this
work is 18 defining the temperature limit of the current line list to be 300~K.
It should be noted that TROVE was also used in the study of the thermal
averaging properties of the spin-spin coupling constants of \nhhh\ by
\citet{10YaYuPa.NH3} and a high-temperature partition function for $^{14}$NH\3\
\cite{14SoHeYu.PH3}.


The paper is structured as follows. In Section~\ref{s:theory} we outline the theoretical approach used for the line
list production. In Section~\ref{s:linelist} the structure of the line list and the description of the quantum numbers
are presented, where some comparisons with experimental data are also given and the accuracy of the line list is
discussed. In Section~\ref{s:concl} some conclusions are offered.

\section{Theoretical approach}
\label{s:theory}

We use the same computational procedure and the associated program TROVE
\cite{07YuThJe.method} as was employed to generate the hot ammonia line list BYTe
\cite{11YuBaTe.NH3}, therefore the reader should refer to this paper for a detailed
description. Here we present only a short outline of this approach.

In order to obtain energies and associated wavefunctions required for building
the line list of \nhhh\ we solve the Schr\"{o}dinger equation for the nuclear
motion variationally. Both the kinetic and potential energy terms of the
Hamiltonian  were expanded to 6th and 8th orders, respectively, in terms of
five linearized coordinates around the reference geometry, defined as a
non-rigid reference configuration associated with the inversion motion
characterized by a relatively low barrier to the planarity. The linearized
coordinates are chosen to be close to the three stretching modes associated
with the N-H vibrations and two asymmetric bending modes combined from the
three bending vibrations of the interbond angles H--N--H. Our vibrational basis
set is a product of six one-dimensional (1D) basis functions. The stretching, bending,
and inversion 1D basis sets are obtained by solving the corresponding reduced
1D Schr\"{o}dinger equations using the Numerov-Cooley
approach~\cite{23Nuxxxx.method,61Coxxxx.method} for each degree of freedom
independently.  This so-called primitive basis set is then
improved through a number of pre-diagonalizations and consecutive contractions.
The latter is controlled by the polyad number
\begin{equation}
P =  2 (v_1+v_2+v_3) + v_4 + v_5 + v_6/2,
\end{equation}
where $v_1, v_2, v_3$ are the quantum numbers associated with the three
stretching modes, $v_4, v_5$ are associated with the asymmetric bending modes,
and $v_6$ counts the inversion mode functions. As in Ref.~\cite{11YuBaTe.NH3},
we define the size of the basis set using the condition  $P \le 14$. We use the
so-called $J=0$ representation, where the final contracted ro-vibrational basis
functions are represented by direct symmetrized products of the vibrational
$J=0$ eigenfunctions and the rigid rotor wavefunctions $| J, K, \tau_{\rm rot}
\rangle$, where $J$ is the rotational angular momentum, $K$ is the projection
of the rotational angular momentum to the molecular axis $z$, and $\tau_{\rm
rot}$ is the rotational parity (see \cite{05YuCaJe.NH3} for further details).
The $J=0$ eigenfunctions are the eigensolutions of the pure vibrational
problem. The highest rotational excitation presently considered is $J=18$. We
only compute and store the energy term values and wavefunctions below 14\,000
\cm\ above the zero point energy (ZPE) obtained as 7414.08~\cm. These
thresholds are chosen to get a reasonable population at room temperature
according with the Boltzmann distribution.

As in \cite{09YuBaYa.NH3} here we employ the EBSC (empirical basis set
correction) scheme, where some of the $J=0$ band centers are substituted with
the corresponding experimental values, where available. For \nhhh\ however
there are only very few band centers known experimentally with high enough
accuracy, namely for $\nu_1$, $\nu_2$, $\nu_3$, $\nu_4$,  $\nu_1+\nu_2$,
$\nu_1+\nu_3$,  $2\nu_2$, $\nu_2+\nu_3$, $2\nu_4^{l=0}$, $2\nu_4^{l=2}$,
$\nu_{1}+2\nu_4$, $\nu_2+2\nu_4$, $2\nu_2+\nu_3$, $\nu_3 + 2\nu_4$ as well as
the ground state inversion splitting
\cite{80Coxxxx.15NH3,82SaHaAm.15NH3,82DiFuTr.15NH3,83UrPaBe.15NH3,83ShScxx.15NH3,83JoPaDc.15NH3,84UrDcRa.15NH3,85UrMiRa.15NH3,85DcUrRa.15NH3,85UrDcRa.15NH3,86SaScxx.15NH3,86UrDcMa.15NH3,87Dcxxxx.15NH3,90FuBaxx.15NH3,06LeLiLi.15NH3,08LeLiXu.15NH3,07LiLeXu.15NH3,11FuNiSp.15NH3,14CeHoVe.15NH3}.
 Therefore the effect of this otherwise
very efficient procedure is rather limited. With this approach the $J=0$
energies are reproduced exactly, while the ro-vibrational coupling leads to a
gradual `de-focus' of the $J>0$ energies. In Table~\ref{t:EBSC} we compare the
original theoretical term values with the experimental band centres used in our
EBSC approach. The Obs.-Calc. residuals in this table  illustrate the
deficiency of our model based on the $^{14}$NH\3\ PES applied for the 15th
isotopologue. Although the accuracy of these particular bands is recovered
through the EBSC approach, the error of other band centers can be expected to
be as large as up to about 0.2~\cm\ at least, as illustrated in Table~\ref{t:EBSC}.


The same PES and DMS as in \cite{11YuBaTe.NH3} were used. The potential energy surface
NH3-Y2010  was obtained by \citet{11YuBaTe.method} by fitting to the experimentally
derived term values of the main isotopologues only, with $J\le 8$ covering term values up
to $E=10\,300$~\cm. Because of the approximations used in the fitting, this
`spectroscopic' PES is an effective object. Therefore it does not guarantee, at least in
principle, the same accuracy for \nhhh\ as was reached for $^{14}$NH\3. We make a
comparison with the experiment in the next section. The \ai\ dipole moment surface ATZfc
DMS used here was developed by \citet{05YuCaLi.NH3} which should be capable of accurate
modelling of \nhhh\ spectra. For the description of the intensity calculations see
Refs.~\cite{05YuThCa.method,11YuBaTe.NH3}.


\begin{figure}[t!]
\centering
\epsfxsize=12.0cm \epsfbox{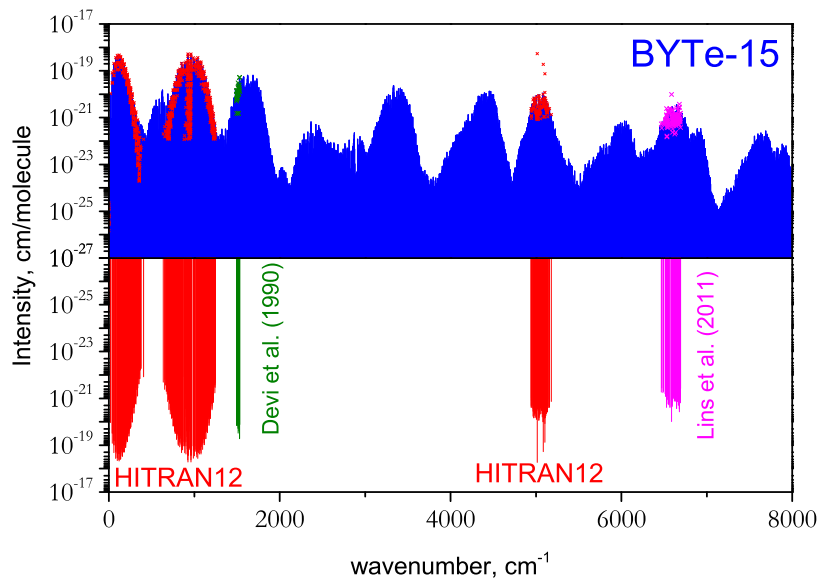}
\caption{Absorption of $^{15}$NH$_3$ at $T=296$~K (log-scale):
The theoretical (BYTe-15 in the upper display) \textit{vs.} experimental line intensities (bottom) from HITRAN~2012,
by \citet{90DeRaPr.15NH3}, and \citet{11LiPfEn.15NH3}. The experimental data points are repeated in the
upper display as crosses for a better illustration of the agreement between theoretical and experimental intensities, where
a number of outliers in the HITRAN data set is also clearly visible.
} \label{f:overview}
\end{figure}

\begin{figure}[t!]
\centering
\epsfxsize=12.0cm \epsfbox{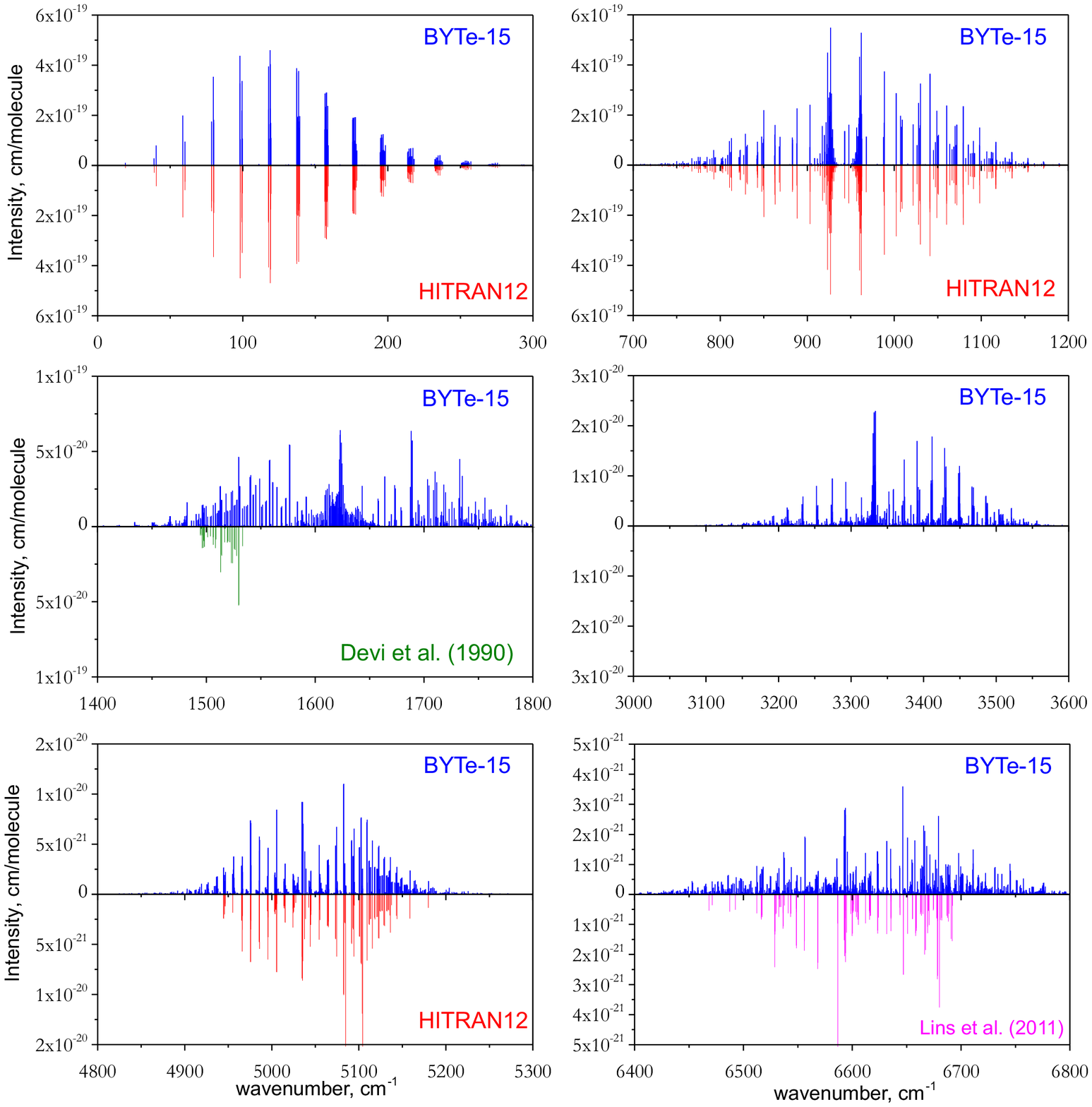}
\caption{Absorption of $^{15}$NH\3\ at $T=296$~K: The theoretical (BYTe-15) vs. experimental line intensities (bottom)
from HITRAN~2012, by \citet{90DeRaPr.15NH3}, and Lins \cite{11LiPfEn.15NH3}. The intensities of the strong 3~$\mu$m band, which is also
shown, are not known experimentally.} \label{f:spectra}
\end{figure}

In Tables~\ref{t:obs-calc1}--\ref{t:obs-calc4} we compare our theoretical term values of
\nhhh\ with their `experimental' counterparts for the vibrational ($J=0$)   and pure
rotational ($J=0\ldots 4 $) states available in the literature (see Introduction). As far
as the accuracy of these term values is concerned it is comparable to the accuracy of the
vibrational term values for the main isotopologue using the same PES. The pure rotational
and rotation-inversional term values  also show a very good agreement with experiment.
This is reassuring especially if the underlying PES was generated using the main
isotopologue only, although  the effect from the isotopic substitution 14 $\to$ 15 is not
expected to be large.

In Table~\ref{t:BYTe-HSL-II} some vibrational term values ($J=0$) of \nhhh\
from this work are compared to the theoretical values computed by
\citet{11HuScLe2.NH3} using their empirical PES HSL-II. The agreement at lower
energies is very good but deteriorates at about 5000~\cm. It is difficult to
claim the better accuracy for any of these two approaches based on this
comparison only. We believe that at least some of our band centers above
6000~\cm\ should be more accurate, see e.g. Table~\ref{t:obs-calc4}. However
according to \citet{14CeHoVe.15NH3} the line positions of \nhhh\ reported by
\citet{11HuScLe2.NH3} are more precise at least for the 2.3~$\mu$m region.

We have compared our intensities to the HITRAN data \cite{HITRAN2012} as well
as to those reported by \citet{90DeRaPr.15NH3} ($\nu_4$) and
\citet{11LiPfEn.15NH3} (near infrared). In Figs.~\ref{f:overview} and
\ref{f:spectra} we show a generated absorption spectrum of \nhhh\ at $T=296$~K
compared to the HITRAN intensities. The agreement is similar to that achieved
by BYTe for $^{14}$NH\3\ \cite{11YuBaTe.NH3}. The \nhhh\ experimental data  is
rather sparse compared to the data available for the main isotopologue. A
number of obvious  outliers (5014.4776,
5084.8734, 5104.2963, and possibly 5103.8909~\cm) in the experimental spectra
indicate  problems with the assignment of the \nhhh\ transitions in
HITRAN. Similar problems have recently been studied and resolved for the
$^{14}$NH\3\ data \cite{13DoHiYu.NH3}. Another  outlier is at
6586.747~\cm\ from the recent work by \citet{11LiPfEn.15NH3} which also appears to
be too strong, see Figs.~\ref{f:overview} and \ref{f:spectra}.


\section{The line list}
\label{s:linelist}

Our room temperature  \nhhh\ line list contains 80\,515\,767 transitions representing
all non-zero ($T=300$~K) intensities covering the wavenumber range up to 8\,000~\cm\
constructed from 270\,646 upper state term values below 14\,000
\cm\ and 9772 lower state term values below 6\,000~\cm\ with rotational excitations up to $J=18$.
Following
Refs.~\cite{06BaTeHa.H2O,11YuBaTe.NH3} we use the two-files ExoMol format
\cite{13TeHiYu.method} to organize the line list for \nhhh. The Energy file
(see an extract in Table~\ref{t:Energy-file}) contains the energy term values
$\tilde{E}_{i}$ (\cm), quantum numbers both in the local and normal mode
representations. Each energy record is indexed with a running number $i$. These
indexes are then used in the Transition file (see extract in
Table~\ref{t:Transit-file}) to refer to a pair of states $i\p$  and $i\pp$
participating in the transition $i\pp \to i\p$. Apart from these indexes, only
the Einstein coefficient $A(i\p,i\pp)$ is needed to complete the transition
record. With this format the size of the line list is significantly reduced.
The line
list
can be also found via
\url{www.exomol.com}
as a part of the ExoMol project \cite{12TeYuxx.methods}. We also supply a
sample Fortran code to be used together with our line list to simulate
intensities or cross sections \citep{13HiYuTe.method}. In fact the unified
ExoMol-format of the present \nhhh\ line list makes this code useful with any
line lists stored in this format.

The largest expansion coefficients of the ro-vibrational eigenfunctions were
used to assign the corresponding final eigenvalues with the vibrational quantum
numbers $v_1, v_2, v_3, v_4, v_5, v_6$, the rotational quantum numbers $J$ and
$K$, the total symmetry $\Gamma_{\rm tot}$ as well as the symmetry of the $J=0$
vibrational basis function $\Gamma_{\rm vib}$. Here $\Gamma_{\rm tot}$  and
$\Gamma_{\rm vib}$ are represented by six irreducible representations $A_1\p$,
$A_2\p$, $E\p$, $A_1\pp$, $A_2\pp$, $E\pp$ in the molecular symmetry group
$D_{3\rm h}$(M) \cite{98BuJexx.method}. In this case our `local' mode basis
functions functions are used as reference and provide approximate labels for the
eigenstates. The problem with this approach (as well as many other assigning
approaches) is the strong mixing of basis set functions at high excitations
which gives rise to the ambiguity in assignment. As a manifestation of the
quality of the assignment we also provide values of the corresponding largest
expansion coefficients, see the $|C_i|^2$-column in Table~\ref{t:Energy-file}:
small numbers (less than 0.5) indicate strong mixing of reference states and show that
that the suggested quantum numbers can be ambiguous.

Recognizing the importance of the conventional `normal' mode quantum numbers, we
map our `local' modes to the `normal' mode quantum numbers using the same
procedure as in Ref.~\cite{11YuBaTe.NH3}. It should be noted however that there
is no direct transformation between these two labelling schemes. Furthermore
due to the approximate nature of the assignment in some cases we obtain
ambiguous normal mode quantum numbers, which do not always correspond to the
experimental normal mode labels. Again, the value $|C_i|^2$ can be used as
measure of this ambiguity.

We follow Ref.~\cite{13DoHiYu.NH3} and define the normal mode quantum numbers
as given by
\begin{equation}
\label{e:quanta}
  n_1, n_2, n_3, n_4, {L_3}, {L_4}, L, \Gamma_{\rm vib}, J, K, i,\Gamma_{\rm rot}, \Gamma_{\rm tot},
\end{equation}
where $L_3= |l_3|, L_4= |l_4|, L= |l|, K = |k|$. Here $n_{i}$, $(i=1,2,3,4)$
are the vibrational normal mode quantum numbers, $l_3$, $l_4$, and $l$ are the
vibrational angular momentum labels; $J$ is the total angular momentum quantum
number, $k = -J,\ldots ,J$ is the projection of the total angular momentum on
the molecule fixed axis $z$; $i=s/a$ is the inversion symmetry of the
vibrational motion; and $\Gamma_{\rm vib}$, $\Gamma_{\rm rot}$ and $\Gamma_{\rm
tot}$ are the symmetry species of the rotational, vibrational, and total
internal wave-functions in the molecular symmetry group $D_{3\rm h}$(M),
respectively, spanning $A_1\p$, $A_1\pp$, $A_2\p$,
$A_2\pp$, $E\p$, and $E\pp$. As was argued by \citet{13DoHiYu.NH3}, the
definition of the signs of the vibrational angular momentum quantum numbers
$l_3, l_4$ and $l$ is ambiguous (as ambiguous the sign of $k$). Therefore we
follow the suggestion of \citet{13DoHiYu.NH3} and use the absolute values $L_i
= |l_i| = n_i, n_i-i , \ldots, 0$ (1) instead.

The symmetries of the initial and final states are important for the line
intensities, which is manifested by the selection rules and
the nuclear statistical weights $g_{\rm ns}$.
Similar to the main isotopologue of ammonia, the ro-vibrational states with the
symmetries of $A_1\p$ and $A_1\pp$ do not exist (i.e. the corresponding $g_{\rm
ns}=0$).  The non-zero nuclear statistical weights factors are 8, 4, 8, 4 for
$A_2\p$, $E\p$, $A_2\pp$, $E\pp$, respectively, which are different from those
of $^{14}$NH\3\ owning to the different nuclear spin of $^{15}$N, 1 against
$1/2$ of $^{14}$N. The TROVE approach uses the symmetrically adapted basis set,
which gives the symmetry labels of the eigenstates automatically. The selection
rules are the following
\begin{equation}
\label{e:rrr}
    A_2\p \leftrightarrow A_2\pp \;, E\p \leftrightarrow E\pp \;
\end{equation}
and
\begin{equation}\label{e:}
    \Delta J = J\p-J\pp = 0, \pm 1\;, \;\; J\p + J\pp \ge 1.
\end{equation}
The non-existing pure vibrational ($J=0$) $A_1\p$ and $A_1\pp$ term values are
also included into the line list with the total statistical weight $g_{\rm tot}
= 0$, which can be useful as band centers.

With our computed energies of \nhhh\ we obtain the partition function of 1165.4
which can be compared to the room-temperature partition function supplied by
HITRAN of 1152.7 by \citet{03FiGaGo.pf}.


\begin{table}
\scriptsize
\caption{\label{t:EBSC}
Comparison of the theoretical term values (\cm) of $^{15}$NH\3\ before EBSC replacement and experimental ones used in the EBSC approach.
See Table~\ref{t:Energy-file} for the description of the notations.
}
\begin{center}
\tabcolsep=5pt
\vspace*{5pt}
\begin{tabular}{cccccrccrrrr@{}l}
    \hline
    \hline
   $\Gamma_{\rm vib}$ &  $v_1$  &  $v_2$   &  $v_3$ &  $v_4$ & $L_3$   &  $L_4$ & $L_4$ & Obs.    &   Calc. &  Obs.-Calc. & \multicolumn{2}{c}{ Ref.} \\
\hline
 $ A_2\pp   $&    0    &  0  &  0  &  0  &  0  &  0  &   0     &        0.761 &         0.758   &$      0.003 $&     \cite{85UrMiRa.15NH3}&      \\
 $ A_1\p    $&    0    &  1  &  0  &  0  &  0  &  0  &   0     &      928.509 &       928.457   &$      0.052 $&     \cite{85UrMiRa.15NH3}&      \\
 $ A_2\pp   $&    0    &  1  &  0  &  0  &  0  &  0  &   0     &      962.912 &       962.894   &$      0.018 $&     \cite{85UrMiRa.15NH3}&      \\
 $ A_1\p    $&    0    &  2  &  0  &  0  &  0  &  0  &   0     &     1591.236 &      1591.185   &$      0.051 $&   \cite{82DiFuTr.15NH3}&$^a$    \\
 $ E\p      $&    0    &  0  &  0  &  0  &  1  &  1  &   1     &     1623.130 &      1623.149   &$     -0.020 $&   \cite{82DiFuTr.15NH3}&$^b$    \\
 $ E\pp     $&    0    &  0  &  0  &  0  &  1  &  1  &   1     &     1624.190 &      1624.202   &$     -0.012 $&   \cite{82DiFuTr.15NH3}&$^b$    \\
 $ A_2\pp   $&    0    &  2  &  0  &  0  &  0  &  0  &   0     &     1870.823 &      1870.853   &$     -0.030 $&     \cite{90FuBaxx.15NH3}&      \\
 $ A_1\p    $&    0    &  3  &  0  &  0  &  0  &  0  &   0     &     2369.274 &      2369.314   &$     -0.041 $&     \cite{82DiFuTr.15NH3}&      \\
 $ E\p      $&    0    &  1  &  0  &  0  &  1  &  1  &   1     &     2533.382 &      2533.380   &$      0.002 $&   \cite{82DiFuTr.15NH3}&$^b$    \\
 $ E\pp     $&    0    &  1  &  0  &  0  &  1  &  1  &   1     &     2577.571 &      2577.590   &$     -0.020 $&   \cite{82DiFuTr.15NH3}&$^b$    \\
 $ A_2\pp   $&    0    &  3  &  0  &  0  &  0  &  0  &   0     &     2876.144 &      2876.130   &$      0.014 $&   \cite{82DiFuTr.15NH3}&$^a$    \\
 $ A_1\p    $&    0    &  0  &  0  &  0  &  2  &  0  &   0     &     3210.614 &      3210.430   &$      0.184 $&     \cite{11FuNiSp.15NH3}&      \\
 $ A_2\pp   $&    0    &  0  &  0  &  0  &  2  &  0  &   0     &     3212.335 &      3212.120   &$      0.215 $&     \cite{11FuNiSp.15NH3}&      \\
 $ E\p      $&    0    &  0  &  0  &  0  &  2  &  2  &   2     &     3234.107 &      3233.925   &$      0.182 $&     \cite{11FuNiSp.15NH3}&      \\
 $ E\pp     $&    0    &  0  &  0  &  0  &  2  &  2  &   2     &     3235.504 &      3235.338   &$      0.165 $&     \cite{11FuNiSp.15NH3}&      \\
 $ A_1\p    $&    1    &  0  &  0  &  0  &  0  &  0  &   0     &     3333.306 &      3333.220   &$      0.086 $&     \cite{11FuNiSp.15NH3}&      \\
 $ A_2\pp   $&    1    &  0  &  0  &  0  &  0  &  0  &   0     &     3334.252 &      3334.160   &$      0.092 $&     \cite{11FuNiSp.15NH3}&      \\
 $ E\p      $&    0    &  0  &  1  &  1  &  0  &  0  &   1     &     3435.167 &      3435.143   &$      0.024 $&     \cite{11FuNiSp.15NH3}&      \\
 $ E\pp     $&    0    &  0  &  1  &  1  &  0  &  0  &   1     &     3435.540 &      3435.475   &$      0.065 $&     \cite{11FuNiSp.15NH3}&      \\
 $ A_1\p    $&    1    &  1  &  0  &  0  &  0  &  0  &   0     &     4288.186 &      4288.024   &$      0.162 $&     \cite{85UrDcRa.15NH3}&      \\
 $ A_2\pp   $&    1    &  1  &  0  &  0  &  0  &  0  &   0     &     4312.345 &      4312.304   &$      0.041 $&     \cite{85UrDcRa.15NH3}&      \\
 $ E\p      $&    1    &  0  &  0  &  0  &  2  &  2  &   2     &     6546.951 &      6546.987   &$     -0.036 $&     \cite{07LiLeXu.15NH3}&      \\
 $ E\pp     $&    1    &  0  &  0  &  0  &  2  &  2  &   2     &     6548.560 &      6548.449   &$      0.111 $&     \cite{07LiLeXu.15NH3}&      \\
 $ E\p      $&    1    &  0  &  1  &  1  &  0  &  0  &   1     &     6596.569 &      6596.605   &$     -0.036 $&     \cite{06LeLiLi.15NH3}&      \\
 $ E\pp     $&    1    &  0  &  1  &  1  &  0  &  0  &   1     &     6597.607 &      6597.498   &$      0.109 $&     \cite{06LeLiLi.15NH3}&      \\
 $ E\p      $&    1    &  0  &  1  &  1  &  0  &  0  &   1     &     6664.486 &      6664.627   &$     -0.141 $&     \cite{08LeLiXu.15NH3}&      \\
 $ E\pp     $&    1    &  0  &  1  &  1  &  0  &  0  &   1     &     6665.480 &      6665.303   &$      0.177 $&     \cite{08LeLiXu.15NH3}&      \\
 \hline
\end{tabular}
\end{center}

$^a$  Estimated from the $a\to s$ band centers.

$^b$ Estimated from the corresponding $^{p}P(J=1,k=1)$ transition frequencies.

\end{table}

\begin{table}
\scriptsize
\caption{\label{t:obs-calc1}
Calculated term values (\cm) of \nhhh\ compared to the experimental values
\citep{85UrDcRa.15NH3}: ground vibrational state.
Here $J$ is the rotational angular momentum, $K$ is its projection, $\Gamma_{\rm tot}$ is the symmetry of the rotational states D$_{3h}$(M), and $s/a$
is the inversion parity.
}
\begin{center}
\tabcolsep=5pt
\vspace*{5pt}
\begin{tabular}{lcccrcrr@{}l}
    \hline
    \hline
 \hline
 $J$  &  $ K$   &  $\Gamma_{\rm tot}$ &    $s/a$ &  Obs.  &   Calc.   &   Obs.-Calc.   \\
 \hline
      0  &         0  &$  A_2^{\prime\prime}          $&     a    &       0.7577  &         0.7577  &$       0.0000 $     \\
      1  &         1  &$  E^{\prime\prime}            $&     s    &      16.1491  &        16.1495  &$      -0.0004 $     \\
      1  &         1  &$  E^{\prime}                  $&     a    &      16.9038  &        16.9042  &$      -0.0004 $     \\
      1  &         0  &$  A_2^{\prime}                $&     s    &      19.8413  &        19.8416  &$      -0.0003 $     \\
      1  &         0  &$  A_1^{\prime\prime}          $&     a    &      20.5892  &        20.5895  &$      -0.0003 $     \\
      2  &         2  &$  E^{\prime}                  $&     s    &      44.7490  &        44.7502  &$      -0.0011 $     \\
      2  &         2  &$  E^{\prime\prime}            $&     a    &      45.5046  &        45.5058  &$      -0.0012 $     \\
      2  &         1  &$  E^{\prime\prime}            $&     s    &      55.8176  &        55.8186  &$      -0.0010 $     \\
      2  &         1  &$  E^{\prime}                  $&     a    &      56.5529  &        56.5540  &$      -0.0011 $     \\
      2  &         0  &$  A_1^{\prime}                $&     s    &      59.5035  &        59.5044  &$      -0.0009 $     \\
      2  &         0  &$  A_2^{\prime\prime}          $&     a    &      60.2322  &        60.2332  &$      -0.0010 $     \\
      3  &         3  &$  A_1^{\prime\prime}          $&     s    &      85.7924  &        85.7948  &$      -0.0024 $     \\
      3  &         3  &$  A_2^{\prime\prime}          $&     s    &      85.7924  &        85.7948  &$      -0.0024 $     \\
      3  &         3  &$  A_2^{\prime}                $&     a    &      86.5526  &        86.5550  &$      -0.0025 $     \\
      3  &         2  &$  E^{\prime}                  $&     s    &     104.2293  &       104.2314  &$      -0.0021 $     \\
      3  &         2  &$  E^{\prime\prime}            $&     a    &     104.9559  &       104.9582  &$      -0.0022 $     \\
      3  &         1  &$  E^{\prime\prime}            $&     s    &     115.2695  &       115.2715  &$      -0.0019 $     \\
      3  &         1  &$  E^{\prime}                  $&     a    &     115.9768  &       115.9789  &$      -0.0021 $     \\
      3  &         0  &$  A_2^{\prime}                $&     s    &     118.9460  &       118.9479  &$      -0.0019 $     \\
      3  &         0  &$  A_1^{\prime\prime}          $&     a    &     119.6469  &       119.6490  &$      -0.0021 $     \\
      4  &         4  &$  E^{\prime}                  $&     s    &     139.2674  &       139.2715  &$      -0.0041 $     \\
      4  &         4  &$  E^{\prime\prime}            $&     a    &     140.0361  &       140.0403  &$      -0.0042 $     \\
      4  &         3  &$  A_2^{\prime\prime}          $&     s    &     165.0675  &       165.0711  &$      -0.0037 $     \\
      4  &         3  &$  A_2^{\prime}                $&     a    &     165.7832  &       165.7931  &$      -0.0099 $     \\
      4  &         2  &$  E^{\prime}                  $&     s    &     183.4416  &       183.4450  &$      -0.0035 $     \\
      4  &         2  &$  E^{\prime\prime}            $&     a    &     184.1315  &       184.1352  &$      -0.0038 $     \\
      4  &         1  &$  E^{\prime\prime}            $&     s    &     194.4443  &       194.4477  &$      -0.0034 $     \\
      4  &         1  &$  E^{\prime}                  $&     a    &     195.1158  &       195.1196  &$      -0.0037 $     \\
      4  &         0  &$  A_1^{\prime}                $&     s    &     198.1083  &       198.1117  &$      -0.0034 $     \\
      4  &         0  &$  A_2^{\prime\prime}          $&     a    &     198.7738  &       198.7775  &$      -0.0037 $     \\
\hline
\hline
\end{tabular}
\end{center}
\end{table}

\begin{table}
\scriptsize
\caption{\label{t:obs-calc2}
Calculated term values (\cm) of \nhhh\ compared to the experimental values
\citep{85UrDcRa.15NH3}: $\nu_2$ state.
Here $J$ is the rotational angular momentum, $K$ is its projection, $\Gamma_{\rm tot}$ is the total symmetry of the ro-vibrational states D$_{3h}$(M), and $s/a$
is the inversion parity.
}
\begin{center}
\tabcolsep=5pt
\vspace*{5pt}
\begin{tabular}{lccccrcrr@{}l}
    \hline
    \hline
 \hline
 $J$  &  $ K$   &  $\Gamma_{\rm tot}$ &  State   &   $s/a$ &  Obs.  &   Calc.   &   Obs.-Calc.   \\
 \hline
      0  &         0  &$  A_1^{\prime}                $&  $\nu_2$              &   s    &      928.457  &        928.457  &$        0.000 $     \\
      1  &         1  &$  E^{\prime\prime}            $&  $\nu_2$              &   s    &      944.594  &        944.590  &$        0.004 $     \\
      1  &         0  &$  A_2^{\prime}                $&  $\nu_2$              &   s    &      948.550  &        948.544  &$        0.006 $     \\
      0  &         0  &$  A_2^{\prime\prime}          $&  $\nu_2$              &   a    &      962.894  &        962.894  &$        0.000 $     \\
      2  &         2  &$  E^{\prime}                  $&  $\nu_2$              &   s    &      972.908  &        972.898  &$        0.010 $     \\
      1  &         1  &$  E^{\prime}                  $&  $\nu_2$              &   a    &      978.924  &        978.922  &$        0.003 $     \\
      2  &         1  &$  E^{\prime\prime}            $&  $\nu_2$              &   s    &      984.763  &        984.749  &$        0.015 $     \\
      2  &         2  &$  E^{\prime\prime}            $&  $\nu_2$              &   a    &     1007.269  &       1007.262  &$        0.007 $     \\
      3  &         3  &$  A_2^{\prime\prime}          $&  $\nu_2$              &   s    &     1013.390  &       1013.371  &$        0.019 $     \\
      2  &         1  &$  E^{\prime}                  $&  $\nu_2$              &   a    &     1018.391  &       1018.382  &$        0.009 $     \\
      2  &         0  &$  A_2^{\prime\prime}          $&  $\nu_2$              &   a    &     1022.096  &       1022.086  &$        0.010 $     \\
      3  &         2  &$  E^{\prime}                  $&  $\nu_2$              &   s    &     1033.137  &       1033.111  &$        0.027 $     \\
      3  &         1  &$  E^{\prime\prime}            $&  $\nu_2$              &   s    &     1044.949  &       1044.918  &$        0.031 $     \\
      3  &         3  &$  A_2^{\prime}                $&  $\nu_2$              &   a    &     1047.921  &       1047.908  &$        0.013 $     \\
      3  &         0  &$  A_2^{\prime}                $&  $\nu_2$              &   s    &     1048.881  &       1048.848  &$        0.032 $     \\
      4  &         4  &$  E^{\prime}                  $&  $\nu_2$              &   s    &     1066.029  &       1065.998  &$        0.030 $     \\
      3  &         2  &$  E^{\prime\prime}            $&  $\nu_2$              &   a    &     1066.450  &       1066.433  &$        0.017 $     \\
      3  &         1  &$  E^{\prime}                  $&  $\nu_2$              &   a    &     1077.544  &       1077.531  &$        0.013 $     \\
      4  &         3  &$  A_2^{\prime\prime}          $&  $\nu_2$              &   s    &     1093.666  &       1093.626  &$        0.040 $     \\
      4  &         4  &$  E^{\prime\prime}            $&  $\nu_2$              &   a    &     1100.870  &       1100.848  &$        0.022 $     \\
      4  &         2  &$  E^{\prime}                  $&  $\nu_2$              &   s    &     1113.317  &       1113.270  &$        0.047 $     \\
      4  &         1  &$  E^{\prime\prime}            $&  $\nu_2$              &   s    &     1125.071  &       1125.020  &$        0.051 $     \\
      4  &         3  &$  A_2^{\prime}                $&  $\nu_2$              &   a    &     1126.801  &       1126.774  &$        0.026 $     \\
      4  &         2  &$  E^{\prime\prime}            $&  $\nu_2$              &   a    &     1145.279  &       1145.249  &$        0.030 $     \\
      4  &         1  &$  E^{\prime}                  $&  $\nu_2$              &   a    &     1156.349  &       1156.317  &$        0.032 $     \\
      4  &         0  &$  A_2^{\prime\prime}          $&  $\nu_2$              &   a    &     1160.036  &       1160.003  &$        0.032 $     \\
\hline
\hline
\end{tabular}
\end{center}
\end{table}

\begin{table}
\scriptsize
\caption{\label{t:obs-calc3}
Calculated term values (\cm) of \nhhh\ compared to the experimental values
\citep{85UrDcRa.15NH3}: The $\nu_1+\nu_2$ band.
Here $J$ is the rotational angular momentum, $K$ is its projection, $\Gamma_{\rm tot}$ is the total symmetry of the ro-vibrational states D$_{3h}$(M), and $s/a$
is the inversion parity.
}
\begin{center}
\tabcolsep=5pt
\vspace*{5pt}
\begin{tabular}{lcccrcrr@{}l}
    \hline
    \hline
 \hline
 $J$  &  $ K$   &  $\Gamma_{\rm tot}$ &    $s/a$ &  Obs.  &   Calc.   &   Obs.-Calc.   \\
 \hline
      0  &         1  &$  A_2\pp                      $&  a           &       4312.303  &    4312.304    &$       -0.001  $      \\
      1  &         1  &$  E\p                         $&  a           &       4328.148  &    4328.151    &$       -0.003  $      \\
      2  &         1  &$  A_2\pp                      $&  a           &       4370.642  &    4370.654    &$       -0.012  $      \\
      2  &         1  &$  E\p                         $&  a           &       4367.042  &    4367.050    &$       -0.008  $      \\
      2  &         1  &$  E\pp                        $&  a           &       4356.231  &    4356.232    &$       -0.001  $      \\
      3  &         1  &$  E\p                         $&  a           &       4425.344  &    4425.364    &$       -0.020  $      \\
      3  &         1  &$  E\pp                        $&  a           &       4414.552  &    4414.563    &$       -0.011  $      \\
      3  &         1  &$  A_2\p                       $&  a           &       4396.541  &    4396.539    &$        0.001  $      \\
      1  &         1  &$  A_2\p                       $&  s           &       4307.725  &    4307.720    &$        0.006  $      \\
      1  &         1  &$  E\pp                        $&  s           &       4303.949  &    4303.944    &$        0.005  $      \\
      2  &         1  &$  E\pp                        $&  s           &       4343.333  &    4343.320    &$        0.013  $      \\
      2  &         1  &$  E\p                         $&  s           &       4332.010  &    4332.002    &$        0.008  $      \\
      3  &         1  &$  A_2\p                       $&  s           &       4406.116  &    4406.086    &$        0.029  $      \\
      3  &         1  &$  E\pp                        $&  s           &       4402.357  &    4402.330    &$        0.028  $      \\
      3  &         1  &$  E\p                         $&  s           &       4391.069  &    4391.046    &$        0.024  $      \\
      3  &         1  &$  A_2\pp                      $&  s           &       4372.198  &    4372.192    &$        0.006  $      \\
      1  &         1  &$  E\p                         $&  a           &       4328.149  &    4328.151    &$       -0.002  $      \\
      2  &         1  &$  E\p                         $&  a           &       4367.042  &    4367.050    &$       -0.009  $      \\
      2  &         1  &$  E\pp                        $&  a           &       4356.230  &    4356.232    &$       -0.002  $      \\
      3  &         1  &$  E\p                         $&  a           &       4425.347  &    4425.364    &$       -0.016  $      \\
      3  &         1  &$  E\pp                        $&  a           &       4414.552  &    4414.563    &$       -0.011  $      \\
      3  &         1  &$  A_2\p                       $&  a           &       4396.543  &    4396.539    &$        0.003  $      \\
      4  &         1  &$  E\p                         $&  a           &       4503.017  &    4503.050    &$       -0.033  $      \\
      4  &         1  &$  E\pp                        $&  a           &       4492.247  &    4492.272    &$       -0.025  $      \\
      4  &         1  &$  A_2\p                       $&  a           &       4474.275  &    4474.288    &$       -0.013  $      \\
      4  &         1  &$  E\pp                        $&  a           &       4449.066  &    4449.061    &$        0.005  $      \\
      1  &         1  &$  E\pp                        $&  s           &       4303.949  &    4303.944    &$        0.005  $      \\
      2  &         1  &$  E\pp                        $&  s           &       4343.336  &    4343.320    &$        0.015  $      \\
      2  &         1  &$  E\p                         $&  s           &       4332.010  &    4332.002    &$        0.007  $      \\
      3  &         1  &$  E\pp                        $&  s           &       4402.358  &    4402.330    &$        0.029  $      \\
      3  &         1  &$  E\p                         $&  s           &       4391.074  &    4391.046    &$        0.028  $      \\
      3  &         1  &$  A_2\pp                      $&  s           &       4372.207  &    4372.192    &$        0.015  $      \\
      4  &         1  &$  E\pp                        $&  s           &       4480.955  &    4480.908    &$        0.047  $      \\
      4  &         1  &$  E\p                         $&  s           &       4469.708  &    4469.668    &$        0.040  $      \\
      4  &         1  &$  A_2\pp                      $&  s           &       4450.918  &    4450.890    &$        0.028  $      \\
      3  &         1  &$  E\p                         $&  a           &       4424.519  &    4425.364    &$       -0.845  $      \\
      2  &         1  &$  A_2\pp                      $&  a           &       4370.642  &    4370.654    &$       -0.012  $      \\
      2  &         1  &$  E\p                         $&  a           &       4367.042  &    4367.050    &$       -0.008  $      \\
      3  &         1  &$  E\p                         $&  a           &       4425.346  &    4425.364    &$       -0.018  $      \\
      3  &         1  &$  E\pp                        $&  a           &       4414.552  &    4414.563    &$       -0.011  $      \\
      0  &         0  &$  A_2\pp                      $&  a           &       4506.607  &    4506.966    &$       -0.359  $      \\
      4  &         1  &$  E\p                         $&  a           &       4503.020  &    4503.050    &$       -0.030  $      \\
      4  &         1  &$  E\pp                        $&  a           &       4492.248  &    4492.272    &$       -0.024  $      \\
      4  &         1  &$  A_2\p                       $&  a           &       4474.268  &    4474.288    &$       -0.020  $      \\
      1  &         1  &$  A_2\p                       $&  s           &       4307.724  &    4307.720    &$        0.005  $      \\
      2  &         1  &$  E\pp                        $&  s           &       4343.335  &    4343.320    &$        0.015  $      \\
      3  &         1  &$  A_2\p                       $&  s           &       4406.115  &    4406.086    &$        0.029  $      \\
      3  &         1  &$  E\pp                        $&  s           &       4402.358  &    4402.330    &$        0.028  $      \\
      3  &         1  &$  E\p                         $&  s           &       4391.068  &    4391.046    &$        0.023  $      \\
      4  &         1  &$  E\pp                        $&  s           &       4480.955  &    4480.908    &$        0.048  $      \\
      4  &         1  &$  E\p                         $&  s           &       4469.711  &    4469.668    &$        0.043  $      \\
      4  &         1  &$  A_2\pp                      $&  s           &       4450.924  &    4450.890    &$        0.033  $      \\
\hline
\hline
\end{tabular}
\end{center}
\end{table}

\begin{table}
\scriptsize
\caption{\label{t:obs-calc4}
Calculated term values (\cm) of \nhhh\ compared to the experimental values
\cite{06LeLiLi.15NH3,07LiLeXu.15NH3,08LeLiXu.15NH3}: the 1.5~$\mu$m band.
Here $J$ is the rotational angular momentum, $K$ is its projection, $\Gamma_{\rm tot}$ is the total symmetry of the ro-vibrational states D$_{3h}$(M), and $s/a$
is the inversion parity.
}
\begin{center}
\tabcolsep=5pt
\vspace*{5pt}
\begin{tabular}{lccccrcrr@{}l}
    \hline
    \hline
 \hline
 $J$  &  $ K$   &  $\Gamma_{\rm tot}$ &  State   &   $s/a$ &  Obs.  &   Calc.   &   Obs.-Calc.   \\
 \hline
      0  &         0  &$  E\p                         $&  $\nu_1+\nu_3$        &   s           &       6596.605  &    6596.569    &$        0.036  $      \\
      0  &         0  &$  E\pp                        $&  $\nu_1+\nu_3$        &   a           &       6597.498  &    6597.607    &$       -0.109  $      \\
      0  &         0  &$  E\p                         $&  $\nu_1+\nu_3$        &   s           &       6664.627  &    6664.627    &$        0.000  $      \\
      0  &         0  &$  E\pp                        $&  $\nu_1+\nu_3$        &   a           &       6665.303  &    6665.303    &$        0.000  $      \\
      1  &         1  &$  E\pp                        $&  $\nu_1+\nu_3$        &   a           &       6612.935  &    6612.745    &$        0.190  $      \\
      1  &         1  &$  E\pp                        $&  $\nu_1+\nu_3$        &   s           &       6613.111  &    6613.311    &$       -0.200  $      \\
      1  &         1  &$  E\p                         $&  $\nu_1+\nu_3$        &   a           &       6613.987  &    6614.222    &$       -0.235  $      \\
      1  &         0  &$  E\p                         $&  $\nu_1+\nu_3$        &   s           &       6616.563  &    6616.585    &$       -0.022  $      \\
      1  &         0  &$  E\pp                        $&  $\nu_1+\nu_3$        &   a           &       6617.389  &    6617.375    &$        0.014  $      \\
      1  &         1  &$  E\pp                        $&  $\nu_1+\nu_3$        &   s           &       6680.448  &    6680.395    &$        0.053  $      \\
      1  &         1  &$  E\p                         $&  $\nu_1+\nu_3$        &   a           &       6681.159  &    6680.969    &$        0.190  $      \\
      1  &         0  &$  E\p                         $&  $\nu_1+\nu_3$        &   s           &       6684.445  &    6684.459    &$       -0.014  $      \\
      1  &         0  &$  E\pp                        $&  $\nu_1+\nu_3$        &   a           &       6685.051  &    6685.062    &$       -0.011  $      \\
      2  &         2  &$  E\pp                        $&  $\nu_1+\nu_3$        &   a           &       6641.024  &    6640.637    &$        0.387  $      \\
      2  &         2  &$  A_2\p                       $&  $\nu_1+\nu_3$        &   s           &       6642.655  &    6642.223    &$        0.432  $      \\
      2  &         1  &$  E\pp                        $&  $\nu_1+\nu_3$        &   s           &       6651.538  &    6651.540    &$       -0.002  $      \\
      2  &         1  &$  E\p                         $&  $\nu_1+\nu_3$        &   s           &       6653.082  &    6653.098    &$       -0.016  $      \\
      2  &         1  &$  E\pp                        $&  $\nu_1+\nu_3$        &   s           &       6653.864  &    6653.364    &$        0.500  $      \\
      2  &         0  &$  E\p                         $&  $\nu_1+\nu_3$        &   s           &       6656.415  &    6656.516    &$       -0.101  $      \\
      2  &         0  &$  E\pp                        $&  $\nu_1+\nu_3$        &   a           &       6657.122  &    6657.101    &$        0.021  $      \\
      2  &         2  &$  A_2\p                       $&  $\nu_1+\nu_3$        &   s           &       6709.141  &    6708.980    &$        0.161  $      \\
      2  &         2  &$  A_2\pp                      $&  $\nu_1+\nu_3$        &   a           &       6709.962  &    6709.633    &$        0.329  $      \\
      2  &         1  &$  A_2\pp                      $&  $\nu_1+\nu_3$        &   a           &       6720.105  &    6720.056    &$        0.049  $      \\
      2  &         1  &$  E\p                         $&  $\nu_1+\nu_3$        &   a           &       6720.618  &    6720.493    &$        0.125  $      \\
      2  &         0  &$  E\p                         $&  $\nu_1+\nu_3$        &   s           &       6724.162  &    6724.185    &$       -0.023  $      \\
      2  &         0  &$  E\pp                        $&  $\nu_1+\nu_3$        &   a           &       6724.623  &    6724.659    &$       -0.036  $      \\
      0  &         0  &$  E\p                         $&  $\nu_1+2\nu_4$       &   s           &       6546.987  &    6546.987    &$        0.000  $      \\
      0  &         0  &$  E\pp                        $&  $\nu_1+2\nu_4$       &   a           &       6548.449  &    6548.449    &$        0.000  $      \\
      1  &         1  &$  A_2\p                       $&  $\nu_1+2\nu_4$       &   a           &       6559.895  &    6559.922    &$       -0.027  $      \\
      1  &         0  &$  E\p                         $&  $\nu_1+2\nu_4$       &   s           &       6567.492  &    6567.504    &$       -0.012  $      \\
      1  &         1  &$  E\pp                        $&  $\nu_1+2\nu_4$       &   s           &       6567.904  &    6567.960    &$       -0.056  $      \\
      1  &         0  &$  E\pp                        $&  $\nu_1+2\nu_4$       &   a           &       6568.657  &    6568.730    &$       -0.073  $      \\
      1  &         1  &$  E\p                         $&  $\nu_1+2\nu_4$       &   a           &       6569.257  &    6569.269    &$       -0.012  $      \\
      2  &         2  &$  E\p                         $&  $\nu_1+2\nu_4$       &   s           &       6582.047  &    6582.056    &$       -0.009  $      \\
      2  &         2  &$  E\pp                        $&  $\nu_1+2\nu_4$       &   a           &       6583.439  &    6583.476    &$       -0.037  $      \\
      2  &         1  &$  A_2\pp                      $&  $\nu_1+2\nu_4$       &   s           &       6599.495  &    6599.514    &$       -0.019  $      \\
      2  &         2  &$  A_2\p                       $&  $\nu_1+2\nu_4$       &   s           &       6601.051  &    6600.952    &$        0.099  $      \\
      2  &         2  &$  A_2\pp                      $&  $\nu_1+2\nu_4$       &   a           &       6602.177  &    6602.120    &$        0.057  $      \\
      2  &         0  &$  E\p                         $&  $\nu_1+2\nu_4$       &   s           &       6608.463  &    6608.506    &$       -0.043  $      \\
      2  &         0  &$  E\pp                        $&  $\nu_1+2\nu_4$       &   a           &       6609.322  &    6609.348    &$       -0.026  $      \\
      2  &         0  &$  E\pp                        $&  $\nu_1+2\nu_4$       &   a           &       6609.323  &    6609.348    &$       -0.025  $      \\
      2  &         1  &$  E\p                         $&  $\nu_1+2\nu_4$       &   a           &       6609.784  &    6609.876    &$       -0.092  $      \\
      2  &         2  &$  E\pp                        $&  $\nu_1+2\nu_4$       &   a           &       6640.194  &    6640.470    &$       -0.276  $      \\
      2  &         1  &$  E\p                         $&  $\nu_3+2\nu_4$       &   a           &       6700.714  &    6699.362    &$        1.352  $      \\
      2  &         2  &$  E\p                         $&  $\nu_3+2\nu_4$       &   s           &       6702.047  &    6703.073    &$       -1.026  $      \\
      2  &         1  &$  A_2\p                       $&  $\nu_3+2\nu_4$       &   a           &       6713.529  &    6712.830    &$        0.699  $      \\
\hline
\hline
\end{tabular}
\end{center}
\end{table}


\begin{table}
\scriptsize
\caption{\label{t:BYTe-HSL-II}
Comparison of the theoretical term values (\cm) of $^{15}$NH\3, from this work (BYTe-15) and computed by \citet{11HuScLe2.NH3}.
See Table~\ref{t:Energy-file} for the description of the notations.
}
\begin{center}
\tabcolsep=5pt
\vspace*{5pt}
\begin{tabular}{cccccrccrrr}
    \hline
    \hline
   $\Gamma_{\rm vib}$ &  $v_1$  &  $v_2$   &  $v_3$ &  $l_3$   &  $v_4$ &  $l_4$ & $\tau_{\rm tor}$ &  BYTe-15    &   HSL-2 &  BYTe-HSL  \\
\hline
   $    A_1\p     $&     0 &      0 &      0 &      0 &     0 &      0 &      0 &      0.00  &       0.00  &$      0.00  $\\
   $    A_2\pp    $&     0 &      0 &      0 &      0 &     0 &      0 &      1 &      0.76  &       0.76  &$      0.00  $\\
   $    A_1\p     $&     0 &      1 &      0 &      0 &     0 &      0 &      0 &    928.46  &     928.47  &$     -0.01  $\\
   $    A_2\pp    $&     0 &      1 &      0 &      0 &     0 &      0 &      1 &    962.89  &     962.93  &$     -0.04  $\\
   $    A_1\p     $&     0 &      2 &      0 &      0 &     0 &      0 &      0 &   1591.18  &    1591.18  &$      0.00  $\\
   $    A_2\pp    $&     0 &      2 &      0 &      0 &     0 &      0 &      1 &   1870.85  &    1870.82  &$      0.03  $\\
   $    A_1\p     $&     0 &      3 &      0 &      0 &     0 &      0 &      0 &   2369.31  &    2369.33  &$     -0.02  $\\
   $    A_2\pp    $&     0 &      3 &      0 &      0 &     0 &      0 &      1 &   2876.13  &    2876.12  &$      0.01  $\\
   $    A_1\p     $&     0 &      0 &      0 &      2 &     0 &      0 &      0 &   3210.43  &    3210.51  &$     -0.08  $\\
   $    A_2\pp    $&     0 &      0 &      0 &      2 &     0 &      0 &      1 &   3212.12  &    3212.08  &$      0.04  $\\
   $    A_1\p     $&     1 &      0 &      0 &      0 &     0 &      0 &      0 &   3333.22  &    3333.27  &$     -0.05  $\\
   $    A_2\pp    $&     1 &      0 &      0 &      0 &     0 &      0 &      1 &   3334.16  &    3334.25  &$     -0.09  $\\
   $    A_1\p     $&     0 &      4 &      0 &      0 &     0 &      0 &      0 &   3438.72  &    3438.70  &$      0.02  $\\
   $    A_2\pp    $&     0 &      4 &      0 &      0 &     0 &      0 &      1 &   4034.03  &    4033.67  &$      0.36  $\\
   $    A_1\p     $&     0 &      1 &      0 &      2 &     0 &      0 &      0 &   4105.77  &    4105.95  &$     -0.18  $\\
   $    A_2\pp    $&     0 &      1 &      0 &      2 &     0 &      0 &      1 &   4161.85  &    4161.73  &$      0.12  $\\
   $    A_1\p     $&     1 &      1 &      0 &      0 &     0 &      0 &      0 &   4288.02  &    4288.02  &$      0.00  $\\
   $    A_2\pp    $&     1 &      1 &      0 &      0 &     0 &      0 &      1 &   4312.30  &    4312.29  &$      0.01  $\\
   $    A_1\p     $&     0 &      5 &      0 &      0 &     0 &      0 &      0 &   4662.71  &    4662.23  &$      0.48  $\\
   $    A_1\p     $&     0 &      2 &      0 &      2 &     0 &      0 &      0 &   4740.39  &    4743.57  &$     -3.18  $\\
   $    A_1\p     $&     0 &      0 &      0 &      3 &     0 &      3 &      0 &   4832.84  &    4832.55  &$      0.29  $\\
   $    A_2\pp    $&     0 &      0 &      0 &      3 &     0 &      3 &      1 &   4834.25  &    4834.26  &$     -0.01  $\\
   $    A_1\p     $&     1 &      2 &      0 &      0 &     0 &      0 &      0 &   4995.22  &    4992.68  &$      2.54  $\\
   $    A_1\p     $&     0 &      0 &      1 &      1 &     1 &      1 &      0 &   5058.33  &    5055.98  &$      2.35  $\\
   $    A_2\pp    $&     0 &      0 &      1 &      1 &     1 &      1 &      1 &   5058.75  &    5056.20  &$      2.55  $\\
   $    A_2\pp    $&     0 &      2 &      0 &      2 &     0 &      0 &      1 &   5074.97  &    5076.68  &$     -1.71  $\\
   $    A_2\pp    $&     1 &      2 &      0 &      0 &     0 &      0 &      1 &   5221.82  &    5220.15  &$      1.67  $\\
   $    A_2\pp    $&     0 &      5 &      0 &      0 &     0 &      0 &      1 &   5322.90  &    5322.97  &$     -0.07  $\\
   $    A_1\p     $&     0 &      3 &      0 &      2 &     0 &      0 &      0 &   5579.27  &    5582.89  &$     -3.62  $\\
   $    A_1\p     $&     0 &      1 &      0 &      3 &     0 &      3 &      0 &   5704.65  &    5705.12  &$     -0.47  $\\
   $    A_1\p     $&     1 &      3 &      0 &      0 &     0 &      0 &      0 &   5723.13  &    5721.57  &$      1.56  $\\
   $    A_2\pp    $&     0 &      1 &      0 &      3 &     0 &      3 &      1 &   5773.60  &    5774.00  &$     -0.40  $\\
   $    A_1\p     $&     0 &      6 &      0 &      0 &     0 &      0 &      0 &   6002.36  &    6001.83  &$      0.53  $\\
   $    A_1\p     $&     0 &      1 &      1 &      1 &     1 &      1 &      0 &   6004.37  &    6007.44  &$     -3.07  $\\
   $    A_2\pp    $&     0 &      1 &      1 &      1 &     1 &      1 &      1 &   6030.73  &    6031.88  &$     -1.15  $\\
   $    A_2\pp    $&     0 &      3 &      0 &      2 &     0 &      0 &      1 &   6106.56  &    6109.59  &$     -3.03  $\\
   $    A_2\pp    $&     1 &      3 &      0 &      0 &     0 &      0 &      1 &   6208.93  &    6208.84  &$      0.09  $\\
   $    A_1\p     $&     0 &      2 &      0 &      3 &     0 &      3 &      0 &   6330.11  &    6333.51  &$     -3.40  $\\
   $    A_1\p     $&     0 &      0 &      0 &      4 &     0 &      0 &      0 &   6346.25  &    6343.14  &$      3.11  $\\
   $    A_2\pp    $&     0 &      0 &      0 &      4 &     0 &      0 &      1 &   6350.08  &    6344.21  &$      5.87  $\\
   $    A_1\p     $&     2 &      0 &      0 &      0 &     0 &      0 &      0 &   6506.86  &    6512.69  &$     -5.83  $\\
   $    A_2\pp    $&     2 &      0 &      0 &      0 &     0 &      0 &      1 &   6508.66  &    6514.29  &$     -5.63  $\\
   $    A_1\p     $&     1 &      0 &      0 &      2 &     0 &      0 &      0 &   6595.10  &    6597.04  &$     -1.94  $\\
   $    A_2\pp    $&     1 &      0 &      0 &      2 &     0 &      0 &      1 &   6595.98  &    6597.93  &$     -1.95  $\\
   $    A_1\p     $&     1 &      0 &      0 &      2 &     0 &      0 &      0 &   6634.75  &    6637.61  &$     -2.86  $\\
   $    A_2\pp    $&     1 &      0 &      0 &      2 &     0 &      0 &      1 &   6636.48  &    6638.79  &$     -2.31  $\\
   $    A_1\p     $&     0 &      4 &      0 &      2 &     0 &      0 &      0 &   6680.14  &    6682.30  &$     -2.16  $\\
   $    A_2\pp    $&     0 &      2 &      0 &      3 &     0 &      3 &      1 &   6692.05  &    6694.31  &$     -2.26  $\\
\hline \hline
\end{tabular}
\end{center}
\end{table}

\begin{table}
\caption{\label{t:Energy-file} Extract from the Energy file.}
\scriptsize \tabcolsep=3pt
\renewcommand{\arraystretch}{1.0}
\begin{tabular}{rrcccccccccccccccccccccc}
    \hline
    \hline
     1     & \multicolumn{1}{c}{2}   &  3  &  4  &  5  &  6  &  7  &  8  &  9  & 10  & 11  & 12  & 13  & 14  & 15  & 16  & 17  & 18  & 19  &   20   & 21  & 22  & 23  & 24  \\
    \hline
  $N$    &  \multicolumn{1}{c}{ $\tilde{E}$}   & $g_{\rm tot}$&$J$&$\Gamma_{\rm tot}$&$n_1$&$n_2$&$n_3$&$L_3$ &$n_4$&$L_4$& $L$& $\Gamma_{\rm vib}$& $s/a$ &$J$&$K$&$\Gamma_{\rm rot}$& $|C_i^{2}|^2$  &$v_1$&$v_2$&$v_3$&$v_4$&$v_5$&$v_6$\\
    \hline
      6997 &    21756.187127 &     8 &     0 &     5 &     0 &     0 &     2 &      0 &      9 &      9 &      9 &      5 &      2 &      0 &      0 &     1 &     0.03 &      0 &      1 &      1 &      0 &     9 &    1   \\
      6998 &    21774.591859 &     8 &     0 &     5 &     0 &     0 &     0 &      0 &     13 &      3 &      3 &      5 &      2 &      0 &      0 &     1 &     0.49 &      0 &      0 &      0 &      0 &    13 &    1   \\
      6999 &    21813.705372 &     8 &     0 &     5 &     0 &     0 &     4 &      4 &      5 &      5 &      9 &      5 &      2 &      0 &      0 &     1 &     0.07 &      1 &      1 &      2 &      0 &     5 &    1   \\
      7000 &    21827.357350 &     8 &     0 &     5 &     0 &     0 &     1 &      1 &     11 &      7 &      6 &      5 &      2 &      0 &      0 &     1 &     0.15 &      0 &      1 &      0 &      0 &    11 &    1   \\
      7001 &    21847.709371 &     8 &     0 &     5 &     0 &     0 &     3 &      3 &      7 &      3 &      0 &      5 &      2 &      0 &      0 &     1 &     0.02 &      1 &      2 &      0 &      0 &     7 &    1   \\
      7002 &    21904.095288 &     8 &     0 &     5 &     0 &     0 &     4 &      4 &      5 &      1 &      3 &      5 &      2 &      0 &      0 &     1 &     0.23 &      1 &      1 &      2 &      0 &     5 &    1   \\
      7003 &    21965.241710 &     8 &     0 &     5 &     0 &     0 &     1 &      1 &     11 &      1 &      0 &      5 &      2 &      0 &      0 &     1 &     0.25 &      0 &      1 &      0 &      0 &    11 &    1   \\
      7004 &    21995.383165 &     8 &     0 &     5 &     0 &     0 &     3 &      1 &      7 &      1 &      0 &      5 &      2 &      0 &      0 &     1 &     0.08 &      0 &      2 &      1 &      0 &     7 &    1   \\
      7005 &    22042.709382 &     8 &     0 &     5 &     0 &     0 &     2 &      0 &      9 &      3 &      3 &      5 &      2 &      0 &      0 &     1 &     0.04 &      0 &      1 &      0 &      0 &    10 &    3   \\
      7006 &     1624.202046 &     4 &     0 &     6 &     0 &     0 &     0 &      0 &      1 &      1 &      1 &      6 &      2 &      0 &      0 &     1 &     0.95 &      0 &      0 &      0 &      0 &     1 &    1   \\
      7007 &     2577.590496 &     4 &     0 &     6 &     0 &     1 &     0 &      0 &      1 &      1 &      1 &      6 &      2 &      0 &      0 &     1 &     0.92 &      0 &      0 &      0 &      0 &     1 &    3   \\
      7008 &     3235.338490 &     4 &     0 &     6 &     0 &     0 &     0 &      0 &      2 &      2 &      2 &      6 &      2 &      0 &      0 &     1 &     0.92 &      0 &      0 &      0 &      0 &     2 &    1   \\
      7009 &     3435.474680 &     4 &     0 &     6 &     0 &     0 &     1 &      1 &      0 &      0 &      1 &      6 &      2 &      0 &      0 &     1 &     0.57 &      0 &      1 &      0 &      0 &     0 &    1   \\
      7010 &     3487.312457 &     4 &     0 &     6 &     0 &     2 &     0 &      0 &      1 &      1 &      1 &      6 &      2 &      0 &      0 &     1 &     0.90 &      0 &      0 &      0 &      0 &     1 &    5   \\
      7011 &     4181.269381 &     4 &     0 &     6 &     0 &     1 &     0 &      0 &      2 &      2 &      2 &      6 &      2 &      0 &      0 &     1 &     0.91 &      0 &      0 &      0 &      0 &     2 &    3   \\
      7012 &     4421.488494 &     4 &     0 &     6 &     0 &     1 &     1 &      1 &      0 &      0 &      1 &      6 &      2 &      0 &      0 &     1 &     0.54 &      0 &      1 &      0 &      0 &     0 &    3   \\
      7013 &     4506.966214 &     4 &     0 &     6 &     0 &     3 &     0 &      0 &      1 &      1 &      1 &      6 &      2 &      0 &      0 &     1 &     0.87 &      0 &      0 &      0 &      0 &     1 &    7   \\
      7014 &     4794.204888 &     4 &     0 &     6 &     0 &     0 &     0 &      0 &      3 &      1 &      1 &      6 &      2 &      0 &      0 &     1 &     0.67 &      0 &      0 &      0 &      0 &     3 &    1   \\
      7015 &     4950.851210 &     4 &     0 &     6 &     1 &     0 &     0 &      0 &      1 &      1 &      1 &      6 &      2 &      0 &      0 &     1 &     0.23 &      0 &      1 &      0 &      0 &     1 &    1   \\
      7016 &     5041.943254 &     4 &     0 &     6 &     0 &     0 &     1 &      1 &      1 &      1 &      2 &      6 &      2 &      0 &      0 &     1 &     0.56 &      0 &      1 &      0 &      0 &     1 &    1   \\
      7017 &     5094.587715 &     4 &     0 &     6 &     0 &     2 &     0 &      0 &      2 &      2 &      2 &      6 &      2 &      0 &      0 &     1 &     0.89 &      0 &      0 &      0 &      0 &     2 &    5   \\
      7018 &     5333.136103 &     4 &     0 &     6 &     0 &     2 &     1 &      1 &      0 &      0 &      1 &      6 &      2 &      0 &      0 &     1 &     0.50 &      0 &      1 &      0 &      0 &     0 &    5   \\
      7019 &     5677.155784 &     4 &     0 &     6 &     0 &     4 &     0 &      0 &      1 &      1 &      1 &      6 &      2 &      0 &      0 &     1 &     0.80 &      0 &      0 &      0 &      0 &     1 &    9   \\
      7020 &     5741.023124 &     4 &     0 &     6 &     0 &     1 &     0 &      0 &      3 &      1 &      1 &      6 &      2 &      0 &      0 &     1 &     0.68 &      0 &      0 &      0 &      0 &     3 &    3   \\
      7021 &     5920.791213 &     4 &     0 &     6 &     1 &     1 &     0 &      0 &      1 &      1 &      1 &      6 &      2 &      0 &      0 &     1 &     0.24 &      0 &      1 &      0 &      0 &     1 &    3   \\
      7022 &     6020.013978 &     4 &     0 &     6 &     0 &     1 &     1 &      1 &      1 &      1 &      2 &      6 &      2 &      0 &      0 &     1 &     0.52 &      0 &      1 &      0 &      0 &     1 &    3   \\
      7023 &     6127.858605 &     4 &     0 &     6 &     0 &     3 &     0 &      0 &      2 &      2 &      2 &      6 &      2 &      0 &      0 &     1 &     0.85 &      0 &      0 &      0 &      0 &     2 &    7   \\
      7024 &     6302.666416 &     4 &     0 &     6 &     0 &     3 &     1 &      1 &      0 &      0 &      1 &      6 &      2 &      0 &      0 &     1 &     0.48 &      0 &      1 &      0 &      0 &     0 &    7   \\
      7025 &     6368.587533 &     4 &     0 &     6 &     0 &     0 &     0 &      0 &      4 &      2 &      2 &      6 &      2 &      0 &      0 &     1 &     0.44 &      0 &      0 &      0 &      0 &     4 &    1   \\
      7026 &     6423.454535 &     4 &     0 &     6 &     0 &     0 &     0 &      0 &      4 &      4 &      4 &      6 &      2 &      0 &      0 &     1 &     0.57 &      0 &      0 &      0 &      0 &     4 &    1   \\
      7027 &     6548.449000 &     4 &     0 &     6 &     1 &     0 &     0 &      0 &      2 &      2 &      2 &      6 &      2 &      0 &      0 &     1 &     0.18 &      0 &      1 &      0 &      0 &     2 &    1   \\
      7028 &     6597.498000 &     4 &     0 &     6 &     1 &     0 &     1 &      1 &      0 &      0 &      1 &      6 &      2 &      0 &      0 &     1 &     0.19 &      2 &      0 &      0 &      0 &     0 &    1   \\
\hline
\hline
\end{tabular}
\begin{tabular}{cll}
             Column       &    Notation                 &      \\
\hline
   1 &   $N$              &       Level number (row)    \\
   2 & $\tilde{E}$        &       Term value (in \cm)                           \\
   3 &  $g_{\rm tot}$     &       Total degeneracy   \\
   4 &  $J$               &       Rotational quantum number    \\
   5 &  $\Gamma_{\rm tot}$$^a$&       Total symmetry in $D_{3\rm h}$(M)           \\
   6,7,8,9 &  $n_1-n_4$   &       Normal mode vibrational quantum numbers (see \protect\cite{13DoHiYu.NH3})    \\
   10,11,  &   $L_3, L_4$ &       Vibrational angular momenta quantum  numbers\\
   12      &   $L$        &       Total vibrational angular momentum quantum  number\\
  13 &  $\Gamma_{\rm vib}$$^a$&       Symmetry of the vibrational contribution in $D_{3\rm h}$(M)     \\
  14 &  $s/a$             &       Inversion symmetry of the vibrational motion; (1,2)  are used for $(s,a)$, respectively   \\
  15 &  $J$               &       Rotational quantum number (the same as column 2)                             \\
  16 &  $K$               &       Rotational quantum number, projection of $J$ onto the $z$-axis                \\
  17 &  $\Gamma_{\rm rot}$$^a$& Symmetry of the rotational contribution in $D_{3\rm h}$(M)       \\
  18 & $|C_i^{2}|$ &  Largest coefficient used in the assignment\\
  19-24 &  $v_1-v_6$         &       Local mode vibrational quantum numbers (see \protect\cite{11YuBaTe.NH3}) \\
\hline
\end{tabular}

$^a$ The symmetry labels (1,2,3,4,5,6) are used for  ($A_1\p$, $A_2\p$,
$E\p$,$A_1\pp$, $A_2\pp$, $E\pp$), respectively.

\end{table}

\begin{table}
\caption{\label{t:Transit-file} Extract from the  Transition file.}
\begin{center}
\scriptsize \tabcolsep=5pt
\renewcommand{\arraystretch}{1.0}
\begin{tabular}{rrc}
    \hline
    \hline
         \multicolumn{1}{c}{$i^{\prime}$}  &  \multicolumn{1}{c}{$i^{\prime\prime}$}  & A$_{i^{\prime}\gets i^{\prime\prime}}$ / s$^{-1}$   \\
\hline
      623528   &       445704 &   1.5466e-08  \\
     1737233   &      1846711 &   1.1752e-07  \\
     1334663   &      1430969 &   2.4884e-06  \\
      446023   &       393427 &   1.1053e-01  \\
      623718   &       688990 &   9.7052e-08  \\
      906264   &       688810 &   7.0643e-10  \\
      483622   &       426792 &   1.7703e-02  \\
      483793   &       539604 &   2.1314e-07  \\
      689346   &       623432 &   3.3598e-03  \\
      502670   &       343967 &   4.9180e-03  \\
     1737113   &      2073678 &   2.3521e-06  \\
      118614   &       148621 &   2.7272e-02  \\
      344060   &       502369 &   1.0103e-09  \\
     1958031   &      1846618 &   1.1248e+00  \\
     1529207   &      1631501 &   1.7843e-02  \\
      561064   &       623436 &   7.7197e-03  \\
     1919239   &      2034874 &   1.5173e-09  \\
       21841   &        12171 &   1.1518e-03  \\
\hline\hline
\end{tabular}
\end{center}
\end{table}

\section{Conclusion}
\label{s:concl}

In this paper a new synthetic line list for $^{15}$NH\3\ is presented. This line list
should be applicable for describing absorption of this molecule for temperatures up to
300~K. The \nhhh\ line list has already proven useful for analysis of the experimental
spectra by \citet{14CeHoVe.15NH3} where it was applied for assignment of the 2.3~$\mu$m
VECSEL spectra of \nhhh.

\section*{Acknowledgements}

This work is supported by  ERC Advanced Investigator Project 267219. I thank J. Tennyson
for helpful discussions and suggestions. I also thank P. \v{C}erm\'{a}k and P. Cacciani
for suggestions related to the ammonia line list.

\bibliographystyle{elsarticle-num-names-2}


\end{document}